\documentclass{article}

\usepackage[nonatbib,final]{neurips_2021}

\usepackage[utf8]{inputenc} 
\usepackage[T1]{fontenc}    
\usepackage{hyperref}       
\usepackage{url}            
\usepackage{booktabs}       
\usepackage{amsfonts}       
\usepackage{nicefrac}       
\usepackage{microtype}      
\usepackage{xcolor}         

\usepackage{graphicx}       
\usepackage{biblatex}       
\usepackage{algorithm2e}    
\title{Decreasing Annotation Burden of Pairwise Comparisons with Human-in-the-Loop Sorting: Application in Medical Image Artifact Rating}

\author{%
  Ikbeom~Jang\thanks{Equal Contribution} \\
  Massachusetts General Hospital \\
  Harvard Medical School \\
  \texttt{ikbeom.jang@mgh.harvard.edu} \\
  \And  
  Garrison~Danley\footnotemark[1] \\
  Massachusetts General Hospital \\
  Harvard Medical School \\
  \texttt{gdanley@mgh.harvard.edu}
  \And  
  Ken~Chang \\
  Massachusetts General Hospital \\
  Massachusetts Institute of Technology \\
  \texttt{kenchang@mit.edu}
  \And  
  Jayashree~Kalpathy-Cramer \\
  Massachusetts General Hospital \\
  Harvard Medical School \\
  \texttt{jkalpathy-cramer@mgh.harvard.edu}
}

\begin{document}

\maketitle

\begin{abstract}
Ranking by pairwise comparisons has shown improved reliability over ordinal classification. However, as the annotations of pairwise comparisons scale quadratically, this becomes less practical when the dataset is large. We propose a method for reducing the number of pairwise comparisons required to rank by a quantitative metric, demonstrating the effectiveness of the approach in ranking medical images by image quality in this proof of concept study. Using the medical image annotation software that we developed, we actively subsample pairwise comparisons using a sorting algorithm with a human rater in the loop. We find that this method substantially reduces the number of comparisons required for a full ordinal ranking without compromising inter-rater reliability when compared to pairwise comparisons without sorting.
\end{abstract}

\section{Introduction}

Deep learning has shown potential for a variety of applications. One key task for machine learning algorithms is the ranking of samples by a quantitative property. Ordinal classification and pairwise comparison are two primary methods for manual annotation needed to train these algorithms. In ordinal classification, the more common approach, a human rater is instructed to identify the closest category for each sample. In pairwise comparison, the rater is instead instructed to pick one of two given samples based on prespecified criteria \cite{furnkranz2010preference, saaty2008relative, thurstone1927law, bradley1952rank}.

Classification rating has been used for a number of tasks in the medical image domain, including disease severity annotation and image quality rating \cite{li2020siamese}. One significant limitation of classification annotations is their subjectivity, which can result in poor inter-rater reliability. For example, a recent study by Kalpathy-Cramer et al. \cite{kalpathy2016plus} found that the ordinal classification method resulted in significantly worse agreement than pairwise comparisons when annotating disease severity in retinopathy of prematurity. They found that while raters might disagree on which class a particular medical image falls into, they are much more likely to agree about which one of a pair of images has more severe disease. Disagreement on classification tasks has been observed not only with expert annotators but with crowdsourced data as well \cite{chang2017revolt, schaekermann2018resolvable, brun2010towards}.

Despite its improved reliability, the pairwise comparison method is less frequently used due to the undesirable scaling of annotation burden with respect to the number of images. Specifically, as the number of images goes up, the total number of comparisons between these images increases quadratically \cite{guo2018experimental}. This quickly becomes a problem for even moderately sized datasets. Methods for reducing the number of pairwise comparisons (sampling) have been studied, including active methods which dynamically choose comparisons based on past input \cite{guo2018experimental, guo2019accelerated, jamieson2011active, li2018hybrid}. Sorting algorithms are one well-known method for ranking based on pairwise comparisons, and have been found to be effective in retrospectively sampling comparisons for preference survey data and animated image data \cite{maystre2017just}. Rating by pairwise comparison is a fairly new concept in the medical domain. To our knowledge, there are no studies which have used a sorting algorithm to actively sample comparisons for a human annotator in real time, nor are we aware of any studies which have used pairwise comparisons to measure medical image artifacts.

In this paper, we describe a real-world application of this annotation strategy to subsample pairwise comparisons using a sorting algorithm with a human in the loop. We find that it is possible to produce an ordinal ranking of input images using substantially fewer comparisons without compromising inter-rater reliability. Furthermore, we hereby release our software tool for medical image annotation to the public, which supports the three annotation strategies used in this paper -- classification, exhaustive comparison, and sort comparison -- as well as image visualization.

\section{Methods}

\textbf{Algorithm}: Our method uses a modern, stable sorting algorithm to suggest a pair of samples to compare in real time based on the set of previous annotations. The TimSort algorithm used for sorting in this study has a theoretical time complexity of \(O(n\ log\ n)\) in the worst case \cite{peters2002timsort}, versus `\(n\) choose 2' \(=n(n-1)/2=O(n^2)\) for exhaustive pairwise comparisons. The use of a sorting algorithm for active sampling creates challenges in terms of software implementation; the program can be closed at any time (e.g., if the annotator takes a break), and integration with user interface libraries can be difficult. In order to simplify this process and avoid having to modify the algorithm directly, we instead store the human rater's previous inputs in a database ourselves, and then replay the sorting algorithm on them when needed. When we come across a comparison which is not yet in the database, the sort is interrupted (by raising an exception) and the comparison is presented to the user for annotation. The comparison is then stored along with its annotation in the database, and we restart the sort again to retrieve the next unlabeled comparison (see Algorithm \ref{alg:pseudocode}).

\begin{algorithm}[H]
{
\small
 \KwData{images}
 \KwResult{ranking of the images}
 \textbf{initialize} comparison\_database, next\_comparison \\

 \textbf{function} sort (images, compare\_func)\{...\} \\
 \textbf{function} compare\_func (image1, image2)\{ \\
 \Indp \textbf{if} annotation for image1 \& image2 is in comparison\_database \textbf{then} \\
   \Indp \textbf{return} annotation for image1 \& image2 from saved\_comparisons \\
 \Indm \textbf{else} \\
   \Indp \textbf{set} next\_comparison to image1 \& image2 \\
   \textbf{raise} Exception ComparisonNotFound \\
 \Indm \textbf{end} \\
 \Indm \textbf{end}\} \\
 
 \textbf{while} \\
   \Indp \textbf{try} \\
     \Indp \textbf{run} sort with images and compare\_func \\
     \textbf{break} \\
   \Indm \textbf{catch} Exception ComparisonNotFound \\
     \Indp do nothing \\
   \Indm \textbf{end} \\
   \textbf{get human input} for next\_comparison \\
   \textbf{store} resulting human annotation in comparison\_database \\
 \Indm \textbf{end} \\
}
 \caption{Pseudo code of pairwise comparisons with human-in-the-loop sorting}
\label{alg:pseudocode}
\end{algorithm}

\begin{figure}[htbp]
  \centering
  \includegraphics[width=1\linewidth]{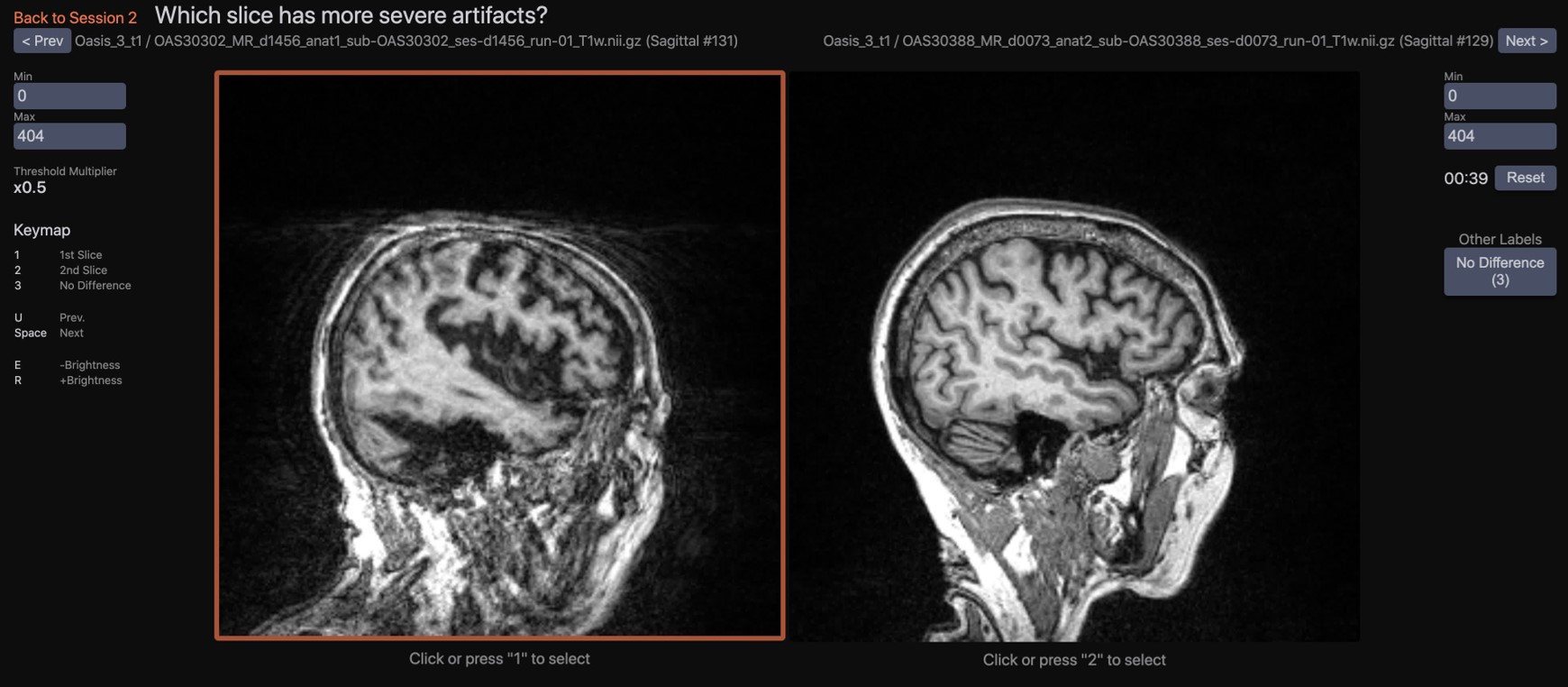}
  \caption{Our annotation software "SliceLabeler" to visualize and grade medical images. All three annotation strategies discussed are available -- classification, exhaustive comparison, and sort comparison.}
\label{fig:software}
\end{figure}

\textbf{Software Tool}: We developed a custom annotation software tool "SliceLabeler" to visualize and annotate medical images, and used this software to compare annotation strategies (See Fig. \ref{fig:software}). Our tool supports all three annotation strategies used in this paper and can record time spent for each annotation. A visualization of the annotation results is shown when annotation is complete and the results can be saved as a CSV file. The software is publicly available at https://github.com/gsnlyd/SliceLabeler.

\textbf{Data}: As a proof of concept, we used the sort-based pairwise comparison method to rank magnetic resonance imaging (MRI) scans of the brain by motion artifact severity. Motion is the most prevalent artifact seen on brain MRI, commonly limiting diagnosis and negatively impacting the radiology workflow, especially when patient recall is required. This initial investigation used fifty T1-weighted sagittal MRI brain slices from randomly selected participants of the Open Access Series of Imaging Studies (OASIS)-3 dataset \cite{lamontagne2019oasis}.

\textbf{Experiments}: Using our tool, two annotators (one technician with 8 years of neuroimaging experience and one technician with 1.5 years of neuroimaging experience) each annotated the same set of data using three different methods: ordinal classification (no, mild, moderate, or severe artifacts; 50 annotations), exhaustive pairwise comparisons (left, right, or no difference; 1225 annotations), and sort-sampled comparisons (left, right, or no difference; 209 annotations on average, see results). The images were presented in the same order to both raters where applicable.
To transform the exhaustive pairwise comparisons into ordinal rankings, a simple counting system was used to generate scores, giving each image +1 for a win, -1 for a loss, and 0 for a tie \cite{shah2017simple, aslam2001models}. The images were then ranked by score. We tried other, more complex rating systems such as Elo \cite{elo1978rating}, but did not see improvements in performance. In sort-sampled comparisons, the stable sorting algorithm TimSort \cite{peters2002timsort} was used to generate pairwise comparisons for the human annotator. When all comparisons are annotated, the final output of the sorting algorithm is an ordinal ranking of the images.

\textbf{Analysis}: We measured four different correlation metrics between the two annotators for all three methods (Ordinal classification, Exhaustive pairwise comparison, and Sort-sampled comparison). Metrics reported are Pearson, Spearman, and Kendall correlations, and Intraclass correlation (ICC; i.e., ICC(2,k)). We also constructed a top-N intersection metric, which we define as the number of images present in the top-N set for both annotators. We measure this top-N intersection metric at every N from 2 to 25 for all three methods to determine whether our sort method is able to find images with severe artifacts with acceptable agreement between annotators.
The ordinal classification results and comparison scores were converted to ordinal rankings by sorting by class and score respectively. To avoid bias from the presentation order, the correlations were computed 100 times, with the images randomized and sorted again each time. This randomization was not used for the sort method, which produces different sets of comparisons for each annotator as the comparisons are sampled actively.
Our software recorded time spent for each annotation. Using this data, we compared per-annotation time and total annotation time between the annotation strategies, while considering annotations that took longer than 20 seconds to be outliers.
Additionally, we measured the number of comparisons required for each method for sets of 30, 50, and 70 images. We compare sort-based sampling to exhaustive comparisons to quantify the improvements in annotation efficiency.

\section{Results}

For 50 images, our sort-sampling method produced an ordinal ranking of the images with 206 and 211 comparisons (comparison count varies per-annotator due to active sampling), as opposed to 1225 comparisons when using conventional exhaustive comparisons. We also measured the number of comparisons required for 30 and 70 images with a single annotator, and found that this reduction is even more substantial with more images. When 30 images were present, 105 sort comparisons were required versus 435 exhaustive comparisons. For 70 images, 310 sort comparisons were required versus 2415 exhaustive comparisons.

\begin{figure}[htbp]
  \centering
  \includegraphics[width=1\linewidth]{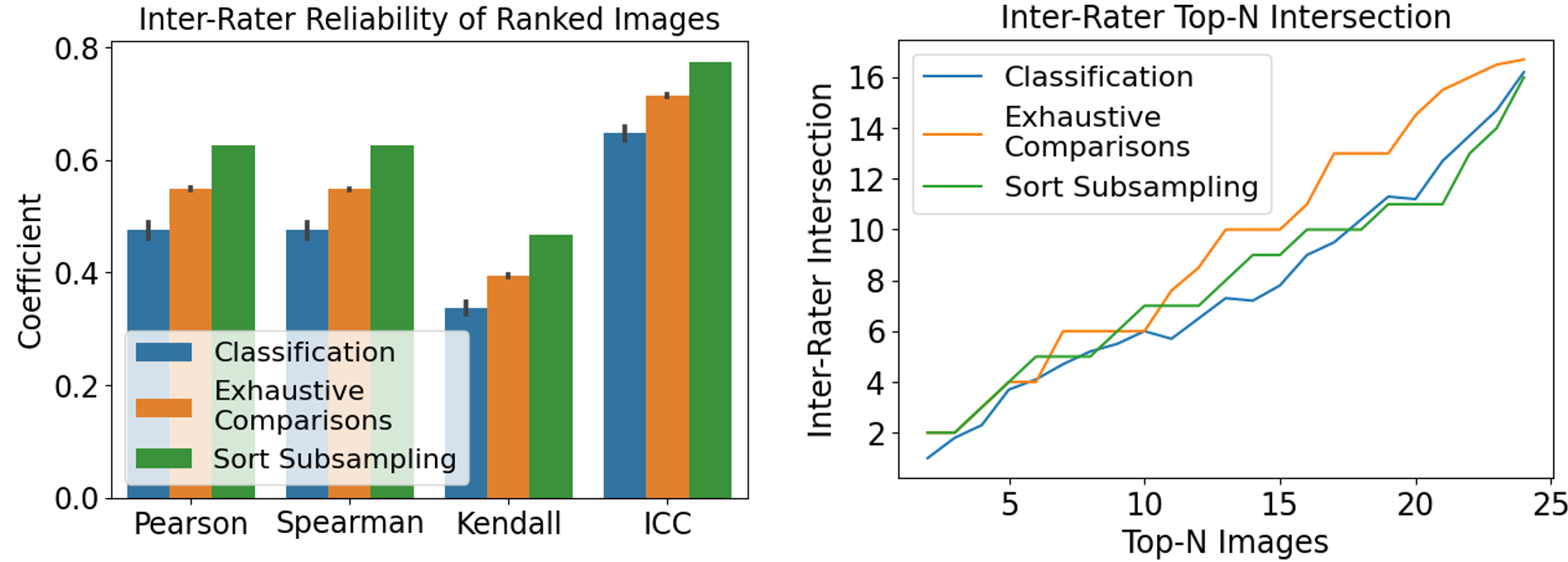}
  \caption{Inter-rater reliability of the three annotation methods: Classification, Exhaustive pairwise comparison, and Sort-sampled pairwise comparison (proposed method).}
\label{fig:reliability}
\end{figure}

We find that despite this huge reduction in the number of comparisons, reliability between raters is not compromised , with correlation scores slightly improved over exhaustive comparisons -- e.g., ICC=0.77 vs 0.71 (\(p<0.0001\)) (See Fig. \ref{fig:reliability}). The exhaustive comparisons produce better correlations than the classification method, and our sort-sampling method achieves slightly better results than exhaustive comparisons while substantially reducing the number of comparisons required.

We observe that our expert annotator spent 2.6 times longer on average (3.17s vs 1.23s; \(p<0.0001\)) to annotate each pair in sort comparison than in exhaustive comparison. Despite this, total annotation time was more than halved by our sorting method (10.63 minutes sort vs 25.10 minutes exhaustive) due to the drastic reduction in the number of comparisons required.

\section{Discussion}

In this paper, we address the high annotation burden of using pairwise comparisons, a more reliable method of ranking samples than conventional ordinal classification. We find that our active sort-sampling method effectively subsamples these comparisons and produces an ordinal ranking without compromising inter-rater reliability. This approach drastically reduced the number of comparisons required, thus decreasing annotation time.

We show that the proposed method is useful for identifying the top-N images in a dataset, e.g. to filter out medical images which are unusable due to severe artifacts. We also believe the method may lead to improved rating quality due to reduced fatigue and pressure from the lower comparison count. This is partially supported by the observation that the annotators spend much less time for each annotation in exhaustive comparison compared to sort comparison, possibly because they were aware of the large annotation workload to be done (1225 for 50 images). On the other hand, our sorting strategy only required 209 annotations on average, and raters spent more time per annotation which could have led to improved rating quality. Additionally, sorting algorithms are well optimized to choose comparisons which are useful for producing a ranking \cite{maystre2017just}, which may have resulted in increased agreement.

For future work, one avenue to explore further is the use of top-K partial sorting algorithms for further efficiency improvements. We also plan to reproduce these results with more annotators and more data in order to further validate our sorting strategy for comparison ranking. Additionally, since we only experimented with one use case (medical image quality), it would be helpful to further evaluate in other areas such as disease severity rating. Finally, it would also be useful to experiment with training neural networks on the generated annotations, as some prior studies \cite{li2020siamese,yildiz2019classification} have found that training on comparison annotations improves neural network performance.

\medskip
{
\small
\printbibliography
}





\end{document}